\documentclass[apj]{emulateapj}
\shorttitle{Core-Collapse Supernovae to $z\approx2.5$ from CANDELS and CLASH}
\shortauthors{Strolger et al.}

\usepackage{natbib}
\usepackage{nicefrac}
\usepackage{ulem}
\usepackage{xcolor}
\usepackage{amssymb,amsmath}
\begin{document}
\title{The Rate of Core Collapse Supernovae to Redshift 2.5 From The CANDELS and CLASH Supernova Surveys}
\author{Louis-Gregory~Strolger\altaffilmark{1}, Tomas~Dahlen\altaffilmark{1,$\dagger$}, Steven A.~Rodney\altaffilmark{2}, Or~Graur\altaffilmark{3,4},  Adam~G.~Riess\altaffilmark{2}, Curtis~McCully\altaffilmark{5,6}, Swara~Ravindranath\altaffilmark{1}, Bahram~Mobasher\altaffilmark{7}, and~A.~Kristin~Shahady\altaffilmark{8}}
\email{strolger@stsci.edu}
\altaffiltext{1}{Space Telescope Science Institute, Baltimore, MD 21218}
\altaffiltext{$\dagger$}{Deceased, February 7, 2014.}
\altaffiltext{2}{Johns Hopkins University, Baltimore, MD 21218}
\altaffiltext{3}{Center for Cosmology and Particle Physics, New York University, New York, NY 10003}
\altaffiltext{4}{American Museum of Natural History, New York, NY 10024}
\altaffiltext{5}{University of California, Santa Barbara, CA 93106}
\altaffiltext{6}{Las Cumbres Observatory Global Telescope Network, Goleta, CA 93117}
\altaffiltext{7}{University of California, Riverside, CA 92521}
\altaffiltext{8}{Florida Institute of Technology, Melbourne, FL 32901}

\begin{abstract}

The Cosmic Assembly Near-infrared Deep Extragalactic Legacy Survey (CANDELS) and Cluster Lensing And Supernova survey with Hubble (CLASH) multi-cycle treasury programs with the {\it Hubble Space Telescope} ({\it HST}) have provided new opportunities to probe the rate of core-collapse supernovae (CCSNe) at high redshift, now extending to $z\approx2.5$. Here we use a sample of approximately 44 CCSNe to determine volumetric rates, $R_{CC}$, in six redshift bins in the range $0.1<z<2.5$. Together with rates from our previous {\it HST} program, and rates from the literature, we trace a more complete history of $R_{CC}(z)$, with $R_{CC}=0.72\pm0.06$ yr$^{-1}$ Mpc$^{-3}$ 10$^{-4}$ $h_{70}^{3}$ at $z<0.08$, and increasing to $3.7^{+3.1}_{-1.6}$ yr$^{-1}$ Mpc$^{-3}$ 10$^{-4}$ $h_{70}^{3}$ to $z\approx2.0$. The statistical precision in each bin is several factors better than than the systematic error, with significant contributions from host extinction, and average peak absolute magnitudes of the assumed luminosity functions for CCSN types. Assuming negligible time delays from stellar formation to explosion, we find these composite CCSN rates to be in excellent agreement with cosmic star formation rate density (SFRs) derived largely from dust-corrected rest-frame UV emission, with a scaling factor of $k=0.0091\pm0.0017\,M^{-1}_{\odot}$, and inconsistent (to $>95\%$ confidence) with SFRs from IR luminous galaxies, or with SFR models that include simple evolution in the initial mass function over time. This scaling factor is expected if the fraction of the IMF contributing to CCSN progenitors is in the 8 to 50 $M_{\odot}$ range. It is not supportive, however, of an upper mass limit for progenitors at $<20\,M_{\odot}$.
\end{abstract}

\section{Introduction}
Core-collapse supernovae (CCSNe) are the explosive end products of massive stars. Despite having similar explosion mechanisms, these events have a wide range of explosion energies due to their broad mass ranges. They are much less useful as cosmological probes, and as such, for nearly two decades, have arguably been a bi-product of large surveys for type Ia supernovae (SNe~Ia). Nonetheless, CCSNe are very much interesting in their own right. They are far better physically understood than their SN~Ia cousins, and most connections to their progenitor stars are far better established.  These supernovae are critical to understanding the formation of dust and the chemical evolution in galaxies and the intergalactic medium. Moreover, as they stem from short-lived stars, CCSN rates virtually trace the rate of instantaneous star formation, providing an independent tracer of star formation in galaxies less sensitive to the ambiguity in extinction corrections that plague far-ultraviolet measures, and free from UV-to-IR light reprocessing assumptions (e.g., contributions from ``warm'' vs. ``cool'' components) inherent in far-infrared measures \citep[cf.][]{Kennicutt:2012fp}. Similarly, tracking CCSN rates over cosmological distances gives a compelling and less biased view of the cosmic star-formation rate density history \citep[cf.][]{Madau:2014fk}.

Until recently, there have not been sufficient samples at significant redshifts to track the CCSN rate history with any precision, largely as SN~Ia surveys tend to selectively bias against CCSNe as they are, on average, a few magnitudes less luminous than SNe~Ia~\citep{Li:2011a}. Still, there have been a small number of complete rate measures at $z > 0.1$, notably from: the Southern Intermediate Redshift ESO Supernova Search~\citep[ $\langle z \rangle = 0.21$]{Botticella:2008}, the Supernova Legacy Survey~\citep[ $\langle z \rangle = 0.3$]{Bazin:2009}, the Subaru Deep Field~\citep[ $\langle z \rangle = 0.66$]{Graur:2011}, the Stockholm VIMOS Supernova Survey~\citep[ $\langle z \rangle = 0.39$ and 0.73]{Melinder:2012}, and the GOODS {\it HST} SN survey~\citep[ $\langle z \rangle = 0.39$, 0.73, and 1.11]{Dahlen:2012}.

The CCSN rates from~\citet[ D12 hereafter]{Dahlen:2012} stand out as the first to extend this rate history to $z>1$. The D12 sample was collected using the {\it HST} Advanced Camera for Surveys (ACS), and used an entirely self-consistent analysis to show unambiguous evolution in CCSN rate history, or $R_{CC}(z)$, extending to $z=1.3$. The D12 results allowed the first viable comparison to consensus cosmic star-formation rate density, or $\psi(z)$, partly derived from rest-trame UV observations of the galaxies in the same Great Observatories Origins Deep Survey (GOODS) fields. While consistent with what would be expected from $\psi(z)$, the D12 $R_{CC}(z)$ measures were dominated by large statistical errors, and nearly equally large systematic biases, the largest of which is attributable to the obscuration (and loss) of events due to line of sight extinction in the SN host.

The two extragalactic programs of the recent multi-cycle treasury program with {\it HST}, the Cosmic Assembly Near-infrared Deep Extragalactic Legacy Survey (CANDELS) and the Cluster Lensing And Supernova survey with Hubble (CLASH), provided new opportunities to probe SN rates in high-$z$ galaxies, both utilizing the Wide Field Camera 3 (WFC3) IR channel, in programs designed to achieve well-resolved ($\sim 0\arcsec.1$ resolution) survey images, to $5\sigma$ sensitivities of $\sim25.5$ mag out to $1.6\mu$m.  These have a distinct advantage over the D12 ACS survey in that they are less sensitive to extinction in host galaxies at $z<1$, and more sensitive to events out to $z=2.5$. Both programs have made some noteworthy discoveries, including GSD10Pri \citep{Rodney:2012}, UDS10Wil \citep{Jones:2013} and GND12Col \citep{Rodney:2014fj}-- three of the highest  redshift SNe~Ia known; EGS11Tyl, GSD12Hum, and GSD12Qua-- three of the highest redshift SNe (of any type) to date \citep{Rodney:2014fj}; and CLO12Car, CLN12Did, and CLA11Tib-- three SNe testing galaxy cluster lens mass models~\citep{Patel:2013}. The analyses on the rates of SNe~Ia from these surveys have been presented in \cite{Rodney:2014fj} and \cite{Graur:2014}.

In this paper we present $R_{CC}(z)$ from the SNe discovered in the CANDELS and CLASH programs, extending CCSN rates for the first time to $z=2.5$, and greatly reducing the statistical errors well below the systematic errors at all lower redshifts. In Section~\ref{sec:survey} we review the observational details of these surveys. In Section~\ref{sec:rates} we describe the selection of approximately 44 of the 94 discovered events we identify as CCSNe, and the calculation of the control-time rates from the sample. In Section~\ref{sec:results} we present the results from this survey, compare them to other rates in the literature and compare the combined CCSNe rate history to the star formation rate density histories as derived from UV and IR studies. Finally, we summarize these results in Section~\ref{sec:summary}. Throughout most of this paper (except where indicated in the error analysis) we assume  $H_0=70$ km s$^{-1}$ Mpc$^{-1}$, $\Omega_M=0.27$, and $\Omega_{\Lambda}=0.73$, consistent with the WMAP7 cosmology~\citep{Komatsu:2011}.

\section{Observational Details}\label{sec:survey}
\subsection{CANDELS}
The CANDELS program surveyed five well-studied deep fields:~GOODS-S, GOODS-N, COSMOS, UDS, and EGS, over a three-year period, each with different total integration times necessary to satisfy ``deep'' and ``wide'' components of the near-IR galaxy survey~\citep{Grogin:2011}. To accommodate searches for supernovae, the observations for each field were broken up into semi-regularly cadenced visits (or epochs), with a total of two visits to each of the ``wide'' fields and 10 to each of the ``deep'' fields, separated by $\sim 50$ days to match SN~Ia rise times at $z\approx1.5$.

The exact details of the SN survey are described in~\citet[ R14 hereafter]{Rodney:2014fj}, but here we provide a review of the survey design. Each single-epoch set of exposures consisted of at least four WFC3/IR detector exposures: two in the $F160W$-passband ($H$), and two in $F125W$-passband ($J)$, each to a $5\sigma$ limiting Vega magnitude of $m_{lim}(H)=25.4$ and $m_{lim}(J)=25.8$, respectively. These passbands were the effective search filters for the SN survey, requiring that events be present at least in both $H$-band exposures, with a modest requirement for visibility in the $J$-band exposures as there is the possibility that $z>2.5$ SNe~Ia could drop out of the bluer passband.

Volumetric SN rate calculations require simulations of each survey field, where in which area, cadence, and depth are critical factors. These simulations can become intensive if each pair of repeated pointings in a mosaic field are considered individually. Generally surveys have a built-in uniformity convenient to considering a few large mosaics over many repeated visits. The CANDELS fields, however, are somewhat less uniform, being composed of nine repeated subareas of very different cadences. 

\begin{figure}[t]
	\centering\includegraphics[width = 3.5in]{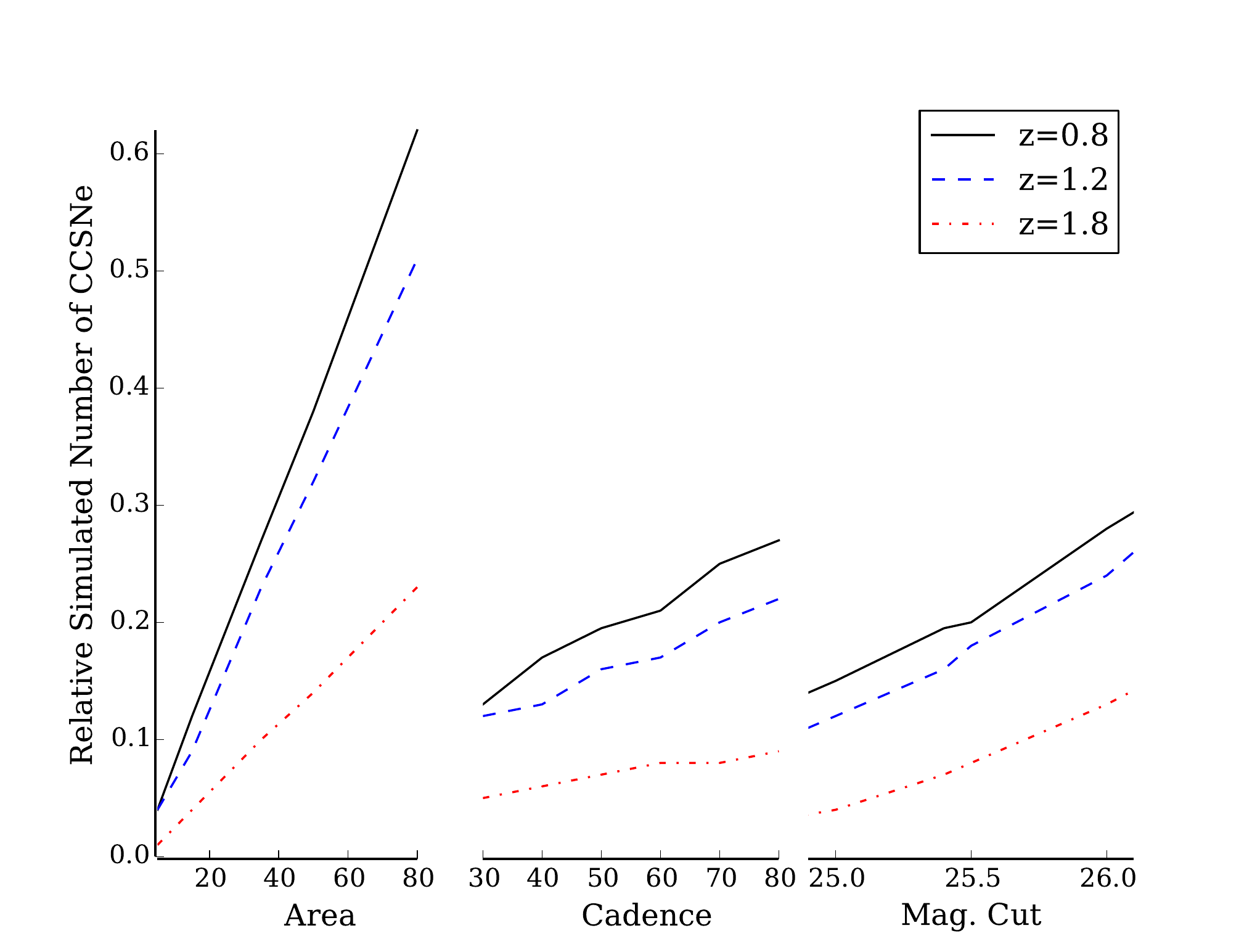}
	\caption{\footnotesize Effect of search area (in arcmin$^2$), cadence (in days), and $F160W$-passband sensitivity (magnitudes) on the simulated number of events at three redshifts. Area has the greatest effect on survey yield, while cadence and depth are secondary factors of the same magnitude.}
	\label{fig:cas}
\end{figure}

One could, as we have done in R14, use an effective area for each subarea, each with a specific average cadence. However, better precision can be obtained using cadence-area groups that better reflect the full range in image combinations. For example, Figure~\ref{fig:cas} shows the effect of varying area, cadence, and depth on the relative expected yield of a mock survey, where area is the principal component, and cadence and depth have similar secondary but not negligible effects. By design, the CANDELS exposure times within a visit are all the same so there is no variance in the survey sensitivity. But selecting an average representative cadence that is 50\% higher or lower than the actual cadence for a field can lead to a 50\% error in the relative number of expected events, if survey area remains unchanged. 

\begin{figure}[t]
	\centering\includegraphics[width = 3.5in]{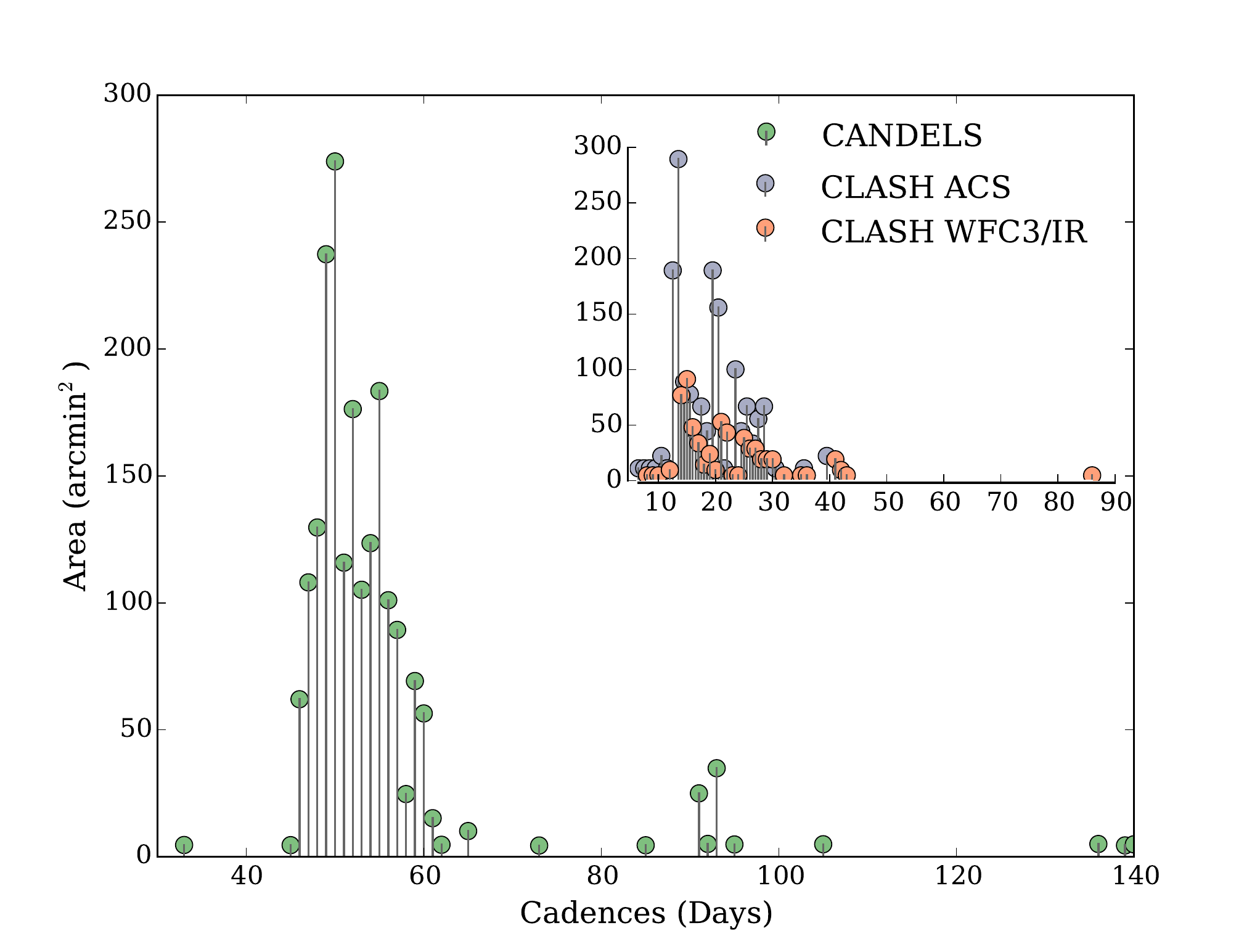}
	\caption{\footnotesize Diagram of the 30 effective areas in CANDELS (green stems), the 24 in CLASH WFC3/IR (red stems), and the 28 in CLASH ACS WFC (blue stems) with similar cadences. The areas and their samples for CANDELS, CLASH WFC3/IR, and CLASH ACS are each treated separately in the rates analysis.}
	\label{fig:cadences}
\end{figure}

For this analysis we collect pointings of similar cadence, regardless of target field, into 30 effective target cadence-areas of a virtual, two-epoch survey. These cadence-areas are shown graphically in Figure~\ref{fig:cadences}, and are broken down by field in the Appendix. Treating the survey in this way has the added benefit of removing a source of confusion in control-time rates (see \S~\ref{sec:rates}) for rolling surveys, where long-term events can be sustained over several epochs of the survey, essentially causing a double-counting problem in the tracking of rising and declining events. 

\subsection{CLASH}
Over the same three-year period, the CLASH program~\citep{Postman:2012} targeted 25 galaxy clusters, with observations spread over up to 16 epochs to enable supernova discovery. The observational strategy allowed for pairs of simultaneous observations with WFC3 and with ACS, with one focused on the target cluster, and the other on one of four parallel positions, constrained by the roll angle of {\it HST}. The $\sim6'$ separation between instruments provided some assurance the deep parallel fields were not significantly influenced by the gravitational lensing from the target cluster, nor contaminated by target cluster members, and therefore essentially equivalent to the ``open'' fields provided by CANDELS. 

As detailed in~\citep[ G14 hereafter]{Graur:2014}, the CLASH parallel fields were essentially surveys of the reddest filters in each detector-set, each visit of which consisting of 2 exposures in WFC3-$H$, and 4 to 6 exposures in the ACS $F850LP$-passband ($Z$), to a $5\sigma$ limiting Vega magnitude of $m_{lim}(H)=25.4$ and $m_{lim}(Z)=25.2$, for up to 4 visits per parallel field. As with CANDELS, simulating the total of 132 $H$ and 148 $Z$-band visits as independent template-search pairs would have been too computationally expensive, so instead we collected each into similar 24 and 28 cadence-area groups for $H$ and $Z$, respectively, as shown in Figure~\ref{fig:cadences} and detailed in the Appendix.

\subsection{Image Processing, SN Detection, and Classification}
SN candidates were discovered in real time, with fast FTP delivery of raw data within a few hours of execution. Our  processing pipeline created searchable difference images within an hour of delivery, the products of which were examined by an experienced search team to select candidates for coordinated followup, usually within 24 hours of the exposures. The processing pipeline has several elements similar to the calibration pipeline used in the OTFR processing by OPUS, providing up-to-date flat field calibrations, up-the-ramp IR sampling with cosmic-ray flagging, and utilizing MultiDrizzle~\citep{Koekemoer:2002} to provide the final sampling to an astrometric grid. The few additional calibrations include improved intra-exposure source-based relative alignment, and modest correction for persistence from exposures within our own program, both of which reduce false detections in the search images. To construct searchable subtraction images, all previous IR imaging in regions that overlap a given incoming epoch were drizzled to the same astrometric grid, and each subtracted from the incoming calibrated data. As there is little variation in the {\it HST} point spread function with WFC3 and ACS, direct subtractions were sufficient without the need to apply image convolution techniques.

SN candidates were identified by human review, and tracked over the regular cadences of the surveys. Additional {\it HST} photometric target-of-opportunity observations were made for many events to gather the most complete multi-wavelength light curves possible for as many candidates as possible, with priority given to candidates likely to be SNe Ia at $z>1$.

Redshift constraints are crucial for developing efficient followup strategies, and the benefit to searching these well-studied fields has been the copious additional information on host galaxies, including existing catalogs with spectroscopic and photometric redshifts~\cite[cf.][]{Dahlen:2013}, and multi-wavelength photometry~\citep{Koekemoer:2011} from which to derive new photometric redshifts when necessary.

All SNe in this sample were photometrically identified using STARDUST based on their observed light curves, and best estimates of their redshifts, photometric or spectroscopic. As discussed in R14, STARDUST uses the SNANA \citep{Kessler:2009a} light curves and SED templates to simulate events of specific types at the redshift of the candidate. Then, through a Bayesian analysis these simulated SNe are compared against the observed data to determine the most likely type and condition (e.g., peak luminosity and extinction) for each candidate event, along with an appropriate range of certainty. While STARDUST is principally designed for identifying SNe~Ia, the Bayesian analysis also provides a measure of a candidate's likelihood of being some other SN type, normalizing such that the sum of probabilities for all SN types is unity. The tool only weighs these possibilities, and assumes the likelihood of other identifications (e.g., AGN, variable stars, or noise) is negligible or completely ruled out by other information, such as a lack of event-host separation, lack of periodicity, or extremely low S/N. It is therefore useful to use the $P_{Ia}$ complement, or  $1-P_{Ia}$, as the probability of a given event being a CCSN.

\section{Determining Volumetric Rates}\label{sec:rates}
We determined the $R_{CC}$ in this analysis using the control-time method, which has some similarities to the monte carlo simulations used by SNANA and other tools, e.g., D12 and R14, but differs in how the effective time of the survey is calculated. The volumetric rate of CCSNe in any redshift interval, represented by a mean or effective redshift, $z$, is expressed as:
\begin{equation}
R_{CC}(z)= \frac{N(z)}{{t_c}(z)\Delta V(z)},
\end{equation}

\noindent where $N(z)$ is the number of observed CCSNe in the redshift interval, and ${t_c}(z)$ is the effective control time in the same redshift interval, or similarly the control time at the effective redshift for the interval. These two terms are discussed further in the subsequent sections. $\Delta V(z)$ is the comoving volume sampled in the redshift interval (i.e., $z_1<z<z_2$), in the solid angle, $\Omega$, of each survey field, such that  $\Delta V(z) = {\Omega}/{4\pi}[V(z_2)-V(z_1)]$.

\subsection{Observed Number of Events}
While in principle it seems easy to count discoveries and sort them by type and redshift, these crucial bits of information are often ambiguous. Few events in our sample have spectroscopic redshifts, determined with certainty from strong features in their hosts' spectra, and fewer still have spectroscopic SN classifications. We rely on photometric redshift techniques for many events, and the STARDUST photometric classifier for most events, to determine probabilistically the most likely scenario of type and redshift. We account for this impreciseness in sample in our rate calculation using an approach described in R14, determining the number of events in any redshift bin, $N(z)$, as the product of the photometric redshift probability distribution function, or $\rm{P}(z|D)$, and the ``CCSN'' classification probability, or $1-P_{Ia}$, for each event, summed for all events, and integrated over the redshift interval. In this way:
\begin{equation}
N(z)\biggl |_{z_1}^{z_2} = \int\limits_{z_1}^{z_2}\sum\limits_{i}^{N} \biggl[\rm{P}_{i}(z'|D)\times\,(1-\rm{P}_{Ia,i})\biggr]\,dz'.\label{eqn:cdf}
\end{equation}
\noindent Events with spectroscopic redshifts have $\rm{P}(z|D)$ that are delta-functions, and those with spectroscopic classifications, SN~Ia or CCSN, have $\rm{P}_{Ia}=1$ or 0, respectively. 

\begin{figure}[t]
	\centering\includegraphics[width = 3.5in]{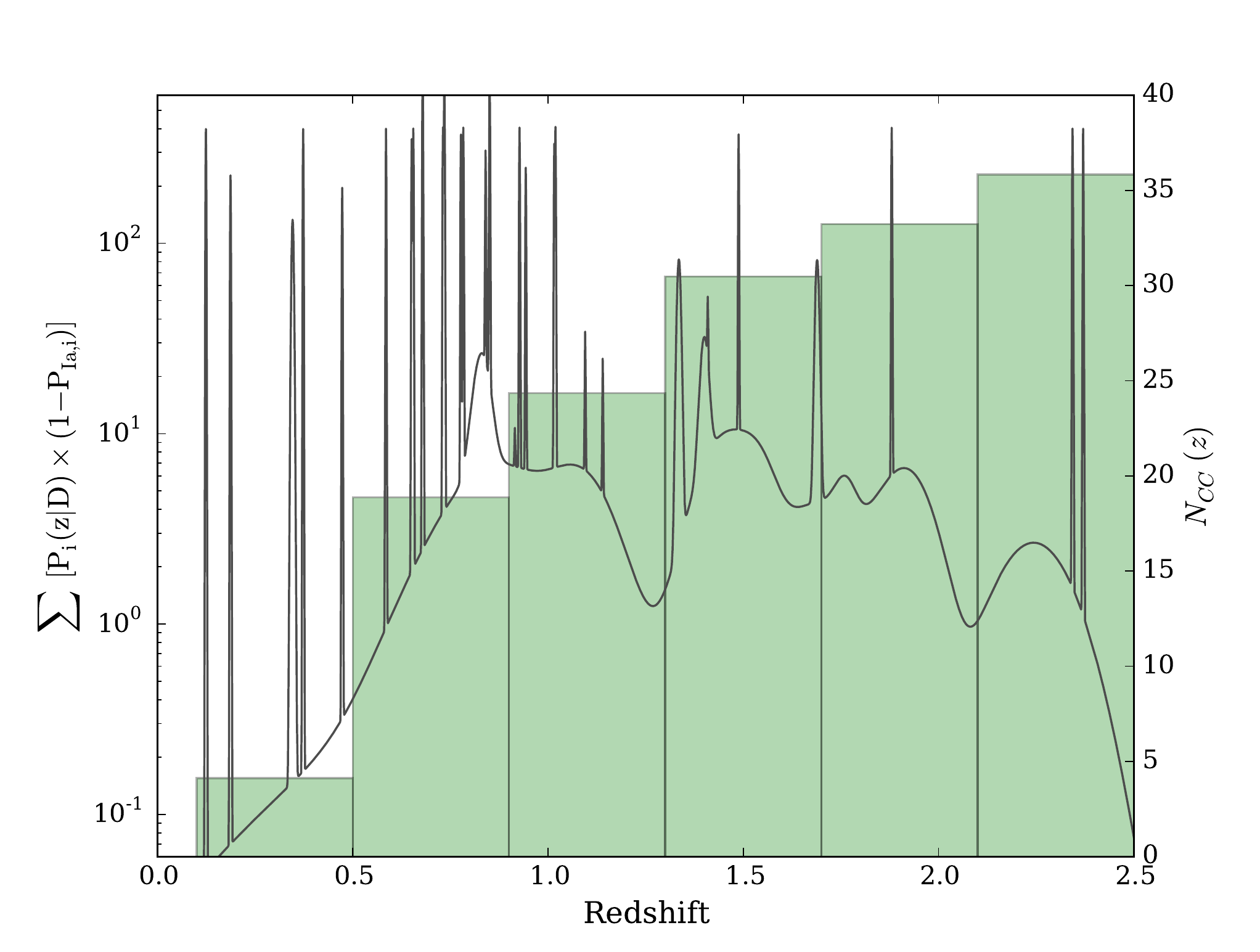}
	\caption{\footnotesize The effective number of CCSNe from the CANDELS survey as a function of redshift. {\it Left ordinate:} the solid line shows the redshift distribution function for all events, determined from $\rm{P}(z|D)$ and CCSN type probabilities, and shown in log space to accentuate the redshift distribution. {\it Right ordinate:} the green region shows the binned cumulative number distribution, or $N(z)$.} 
	\label{fig:number}
\end{figure}

\begin{figure}[t]
	\centering\includegraphics[width = 3.5in]{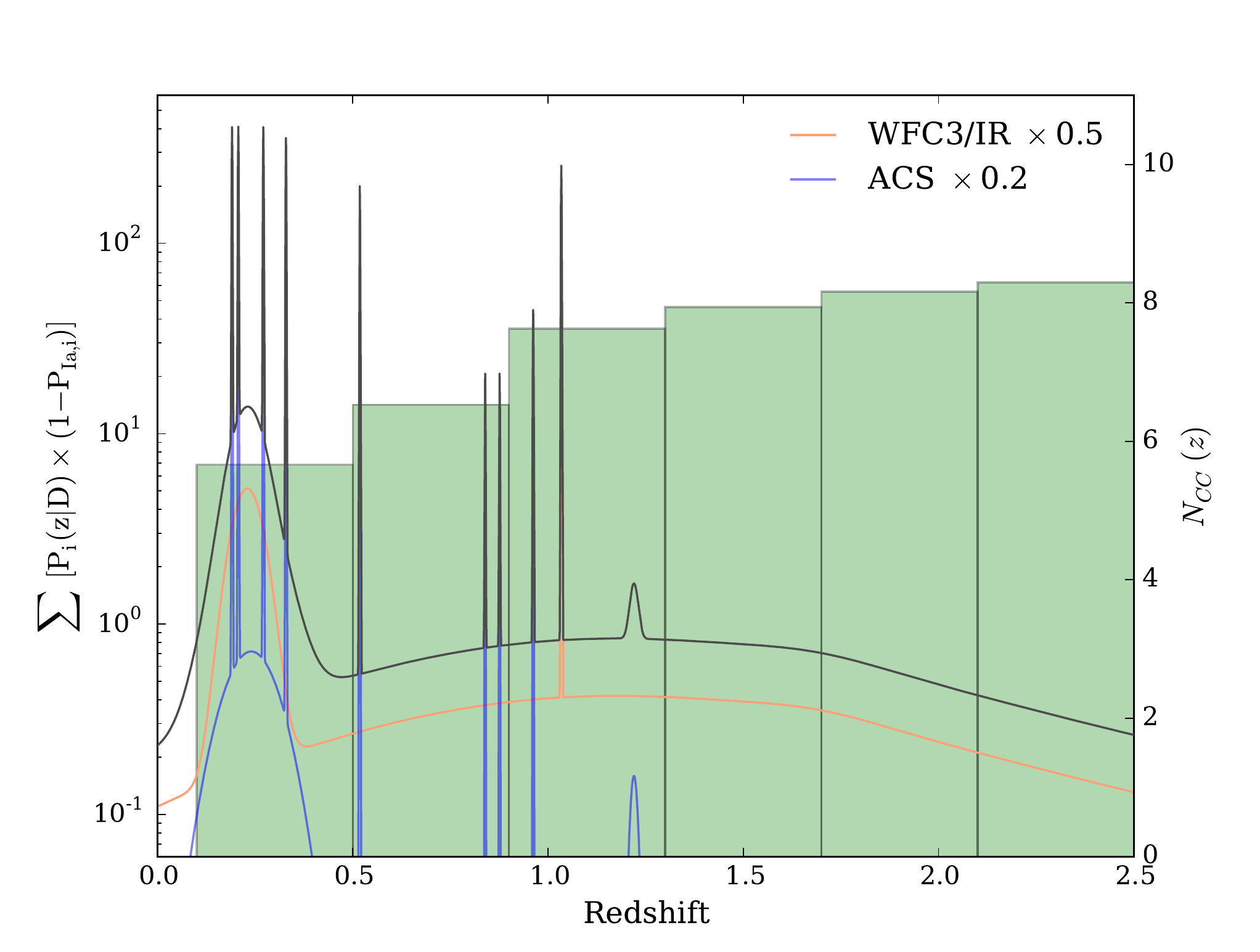}
	\caption{\footnotesize Same as Figure~\ref{fig:number}, the effective number of CCSNe from the CLASH survey as a function of redshift. The blue and orange lines are for the ACS and WFC3/IR parallel fields, respectively, scaled by values in legend to highlight their relative contributions to the total redshift distribution function (in black), and binned cumulative number distribution (in green). }
	\label{fig:number2}
\end{figure}

To allow a more direct comparison to previous rates, we calculate our CCSNe rates in six redshift bins centered at redshifts 0.3, 0.7, 1.1, 1.5, 1.9 and 2.3, each with a bin width of $\Delta z = 0.4$.  Both the cumulative redshift distribution, and the binned cumulative number distribution used in the valuation of the rates are shown in Figures~\ref{fig:number} and \ref{fig:number2}. 

\subsection{The Control Time}
The control time method addresses the probability of detecting a SN of a given type, redshift, peak luminosity, age, and extinction expressed as an amount of time in which the said event would be visible, given the depth, area, and cadence of the survey. It starts in very much the same way that monte carlo methods do, utilizing libraries of light curves and spectra to simulate SNe with assumed conditions, under the constraints of the assumed distributions of each of these conditions. The monte carlo approach, however, samples from these distributions ideally placing simulated events in real search data to test their recovery, often in real time. The control time approach is to first determine survey sensitivities (perhaps in real time), and then iteratively test the recoverability of all model scenarios against this sensitivity threshold, yielding more accurate but computationally intensive results.

The control time ($t_c$) itself is essentially the amount of time our survey is sensitive to CCSNe of a given type at a specified redshift. In an analytical approach, we determine the likelihood of detecting a model CCSN type considered at a light-curve age of $t$ by differencing the model flux at this age from the model flux at an earlier age, given by a time-dilation corrected cadence. We compare this flux difference to our survey efficiency in detecting objects of similar differenced flux, resulting in a likelihood, $\varepsilon$, for detecting the CCSN model at that light-curve age. We integrate over all plausible light-curve ages, from explosion to several hundred days after explosion in the rest frame of the modeled event, giving the total effective time that SN model would be visible within the survey, or,
\begin{equation}
t_c(\beta) =\int\limits_{t=0}^{\infty} \varepsilon(\beta; t)\, dt.
\label{eqn:tca}
\end{equation}
\noindent Here, $\beta$ holds the adjustable parameters of the model (SN type and redshift) and the survey (cadence, filter, and sensitivity). 

The survey efficiency was determined from empirical tests of fake SNe in real data (see R14 and G14 for details). We express it here as
\begin{equation}
\varepsilon (m)=\{1+\exp[0.4(m-m_c)]\}^{-1},
\end{equation}
\noindent where $m$ is the magnitude corresponding to the difference in flux between the detection and template epochs, and $m_c$ is the magnitude cutoff, where transients with $m>m_c$ were recovered $<50\%$ of the time. This value is $m_c(H)=25.4$ for CANDELS and the WFC3/IR component of CLASH, and $m_c(Z)=25.2$ for the ACS/WFC component of CLASH.

Even within a given type, CCSNe have a range of peak luminosities (characterized by $B$-band peak absolute magnitude, $M_B$), and they are likely to experience some extinction internal to their host galaxy, which we characterize by the extinction parameter in the $V$-band, or $A_V$. To account for each of these possibilities (discussed in further detail in later sections), we evaluate both as changes to the observed flux, and determine the expectation of $t_c$ using probability distribution functions, $P(M_B)$ and $P(A_V)$, respectively, such that
\begin{align}
t_c(\beta) = \int\limits_{0}^{\infty} \int\limits_{0}^{\infty} \int\limits_{-\infty}^{\infty} &\varepsilon(\beta; t, M_B, A_V)\, P(M_B)\nonumber\\
&\times P(A_V)\, dM_B\, dA_V\, dt,
\label{eqn:tcb}
\end{align}
thereby defining a control time in the rest frame for an entire CCSN type at that redshift in our survey. As the details of both surveys have been discussed in Section~\ref{sec:survey}, and more extensively in R14 and G14, we will focus the following discussion on the selected CCSN models and their dependencies.

\subsection{CCSN Types}
CCSNe have very different light curve shapes, peak luminosities, and relative rates of occurrence resulting in different visibilities in our survey.  For individual assessment, we separate the CCSNe into five main types: Type II Plateau (IIP), Type II Linear (IIL), Type II with narrow emission lines (IIn), and stripped envelope Types, Ib and Ic. We adopt mean absolute magnitudes and dispersions determined from low-$z$ samples from the Asiago Supernova Catalog~\cite[ASC,][]{Richardson:2014fk}, reproduced here in Table~\ref{tab:subtypes}, and use them as Gaussian-distributed luminosity functions in $P(M_B)$.

\begin{deluxetable}{lccr}
\tabletypesize{\footnotesize}
\tablecolumns{4}
\tablewidth{0pt}
\tablecaption{\sc Subtypes of Core Collapse Supernovae}
\tablehead{\colhead{Type}&\colhead{Peak $M_B$\tablenotemark{a}}&\colhead{Dispersion}&\colhead{Fraction\tablenotemark{b}}
\label{tab:subtypes}}
\startdata
IIP& $-16.80\pm0.37$ &0.97 & 52.4\%\\
IIL& $-17.98\pm0.34$ & 0.90 & 7.3\%\\
IIn& $-18.62\pm0.32$ & 1.48 & 6.4\%\\
Ib& $-17.54\pm0.33$ & 0.94 & 6.9\%\\
Ic& $-16.67\pm0.40$ & 1.04 & 17.6\%\\
Faint CCSNe& $>-15$&\nodata& 9.4\%\\
\enddata
\tablenotetext{a}{\footnotesize From~\cite{Richardson:2014fk}, where uncertainty is the standard error in the mean. Dispersion is the standard deviation of the sample.}
\tablenotetext{b}{\footnotesize From ~\cite{Li:2011a}}
\end{deluxetable}

The ASC sample relied heavily on discoveries as reported by the community, and therefore is likely subject to discovery biases of from varied (and perhaps unknown) survey cadences. In contrast, the Lick Observatory Supernova Search~\cite[LOSS,][]{Li:2011a} had a strategy of systematically repeated visits to the same target galaxies, which suggests a greater completeness in polling CCSN diversity than ASC. However admittedly, ASC may compensate by nature of having a larger volume-limted sample, to 100 Mpc rather than 60 Mpc. We nonetheless adopt the volume-limited, Malmquist-bias corrected fractions of CCSNe for each type from LOSS as the relative frequency from each contributing to the total number of expected events. Those fractions are also shown in Table~\ref{tab:subtypes}.

The SNANA library includes some well studied CCSNe, each with well-sampled $ugriz$ light curves and temporally matched optical (3000~{\AA} to 1~$\mu m$) spectra. We use these light curves and spectra to build mean templates for each of the CCSN types, ultimately constructed from 28 SNe~IIP, 2 SNe~IIn, 11 SNe~Ib, and 8 SNe~Ic from the Sloan Digital Sky Survey and the Carnegie Supernova Project, and a synthetic SN~IIL model (cf.~the SNANA handbook). We normalize these light curves to the average peak $M_B$ for each class from Table~\ref{tab:subtypes}.

For simplicity, we assume the luminosity functions of each CCSN subtype are Gaussian [$N(M_B,\sigma)$], with means and dispersions from Table~\ref{tab:subtypes}, and treat these as probability distributions for Equation~\ref{eqn:tcb} such that $P(M_B)\equiv N(M_B,\sigma)$. We also modify the observed dispersion to account for an additional scatter due to gravitational lensing, which we discuss in Section~\ref{sec:lens}.

As discussed in D12, there are populations of rare CCSNe that are not well accounted for in the literature. For example, there is a population of SN~1987A-like and other less luminous CCSNe ($M_B > -15$) that do not have sufficient numbers for assessment in LOSS or ASC samples. Estimates on the fraction faint CCSNe produced range to as high as 30\%~\citep{Li:2011a}, however, there is not yet a complete sample to adequately characterize the luminosity function of these subtypes in the low-$z$ universe. But, by being intrinsically faint, their contribution to our CCSN rates is expected to be limited to lower redshifts, perhaps contributing only to our lowest redshift bin. Conservatively, we estimate their contribution at $\sim10\%$. 

There are also classes of super-luminous events~\citep{Gal-Yam:2012,Heger:2002qf}, whose origins remain unclear but are likely tied to extremely massive stars, which have the potential to add to every redshift bin. However, as they are exceedingly rare occurrences at low-$z$ (perhaps $<2\%$), we chose not to account for these in our expected samples. We note, however, that some super-luminous events may stem from population III stars, implying that their number could increase with redshift. As this possibility is only weakly explored in the literature, we leave this as a caveat to this study.

\subsection{K-Corrections}
The template light curves are generally not well-matched to the part of the rest-frame spectral energy distributions (SEDs) covered by the $H$ and $Z$-passbands of the surveys. We apply k-corrections to the light curve templates to simulate the flux in this synthetic rest-frame passband. We calculate the k-corrections following the prescription of \cite{Kim:1996},

\begin{align}
k(t) =& -2.5\Biggl[\log\Biggl(\frac{\int H(\lambda[1+z])\, F(\lambda,t)\,d\lambda}{\int S_y(\lambda)\, F(\lambda,t)\, d\lambda}\Biggr)\nonumber\\
&-\log\Biggl(\frac{\int H(\lambda)\, \mathcal{Z}(\lambda)\,d\lambda}{\int S_y(\lambda)\, \mathcal{Z}(\lambda)\, d\lambda}\Biggr)\Biggr],
\end{align}

\noindent where $F(\lambda, t)$ is the spectral energy distribution of the SN at the time of observation,  $H(\lambda)$ is the transmission of the survey passband, $H(\lambda[1+z])$ is a synthetic survey filter blue-shifted to the rest frame of the SN, and $S_y(\lambda)$ is a rest-frame template passband ($ugri$ or $z$) that provides the closest match to the synthetic filter. $ \mathcal{Z}(\lambda)$ is the Vega spectral energy distribution~\citep{Colina:1996qd} that defines the zero-magnitude base for the magnitude system. 

\subsection{Extinction Correction}\label{sec:ext}
For most of this analysis, we consider CCSN hosts as normal galaxies well-represented by the sample of CCSN hosts seen in the local universe. At low-$z$,
the observed distribution of host extinction, $A_V$, for CCSNe is exponential~\citep{Hamuy:2002uq, Schmidt199442}, which is supported by dust modeling of host galaxy extinction effects on the observed rates of CCSNe~\citep{Hatano:1998, Riello:2005}. This extinction distribution is corrected using the \cite{Calzetti:2000vn} extinction law (with $R_V=4.05$) from the rest-frame $V$-band to the effective wavelength of their synthetic rest-frame filter covered by the survey passbands. We evaluate this distribution as the extinction likelihood function, where $P(A_V)\propto\exp(-\lambda_V\,A_V)$ in the control time calculation of Equation~\ref{eqn:tcb}, with $\lambda_V = 0.187$ in the rest frame $V$-band, and normalizing such that $\sum P(A_V)=1$.

\subsection{Gravitational Lensing Degradation of the Luminosity Function}\label{sec:lens}
As light from distant sources propagates to us, it is perturbed by the numerous gravitational potentials of large scale structures along the way. The amplifying and de-amplifying the source as its photons are scattered into or out of our line of sight should result in a net zero effect, on average. However this lensing effect will add significantly to the dispersion in  observed luminosity functions, and add an ever increasing dispersion with redshift. We use the simulated predictions from \cite{Holz:2005fk}, which suggest the observed distribution in absolute magnitudes, given a gaussian-distributed internal dispersion in magnitude space, as a function of redshift will be

\begin{equation}
\sigma_{m}(z)=\sqrt{\sigma_{\rm int}^2+(0.093\,z)^2},
\end{equation}
\noindent where $\sigma_{\rm int}$ is the rest-frame dispersions given in Table~\ref{tab:subtypes}. We use this corrected dispersion in the adopted normally-distributed luminosity functions of Equation~\ref{eqn:tcb}, where $P(M_B)\equiv N[M_B,\sigma_{m}(z)]$. 

\section{Results}\label{sec:results}
\begin{deluxetable*}{lcccccc}
\tabletypesize{\footnotesize}
\tablecolumns{7}
\tablecaption{\sc Volumetric CCSN Rates From CANDELS and CLASH}
\tablehead{\colhead{}&\multicolumn{2}{c}{CANDELS}&\multicolumn{2}{c}{CLASH}&\multicolumn{2}{c}{CANDELS+CLASH}\\
\colhead{Redshift}&\colhead{Rate\tablenotemark{a}}&\colhead{N$_{CC}$\tablenotemark{b}}&\colhead{Rate}&\colhead{N$_{CC}$}&\colhead{Rate\tablenotemark{c}}&\colhead{N$_{CC}$}\label{tab:rates}}
\startdata
0.3$\pm0.2$ & $1.34^{+1.03}_{-0.60}$ & 4.1 & $3.61^{+2.68}_{-1.64}$ & 5.7 & $1.97^{+1.45}_{-0.85}$ & 9.8 \\
0.7$\pm0.2$ & $4.61^{+1.52}_{-1.20}$ & 14.8 & $0.55^{+2.09}_{-0.45}$ & 0.9 & $2.68^{+1.54}_{-1.04}$ & 15.6 \\
1.1$\pm0.2$ & $1.73^{+1.09}_{-0.70}$ & 5.5 & $1.49^{+7.72}_{-0.78}$ & 1.1 & $1.70^{+1.19}_{-0.71}$ & 6.6 \\
1.5$\pm0.2$ & $3.38^{+1.98}_{-1.30}$ & 6.1 & $2.41^{+15.3}_{-1.99}$ & 0.3 & $3.25^{+2.03}_{-1.32}$ & 6.4 \\
1.9$\pm0.2$ & $3.15^{+3.24}_{-1.73}$ & 2.8 & $3.21^{+28.1}_{-2.66}$ & 0.2 & $3.16^{+3.37}_{-1.77}$ & 3.0 \\
2.3$\pm0.2$ & $6.15^{+6.59}_{-3.47}$ & 2.6 & $6.41^{+91.2}_{-5.30}$ & 0.1 & $6.17^{+6.76}_{-3.52}$ & 2.7 \\
\enddata
\tablenotetext{a}{\footnotesize In units yr$^{-1}$ Mpc$^{-3}$ 10$^{-4}$ $h_{70}^{3}$ with statistical errors.}
\tablenotetext{b}{\footnotesize From Equation~\ref{eqn:cdf}, in the intervals specified in the redshift column.}
\tablenotetext{c}{\footnotesize Weighted averages of CCSNe rates. Uncertainties are standard errors in the weighted means.}
\end{deluxetable*}

\begin{figure*}[t]
	\centering\includegraphics[width = 5.25in]{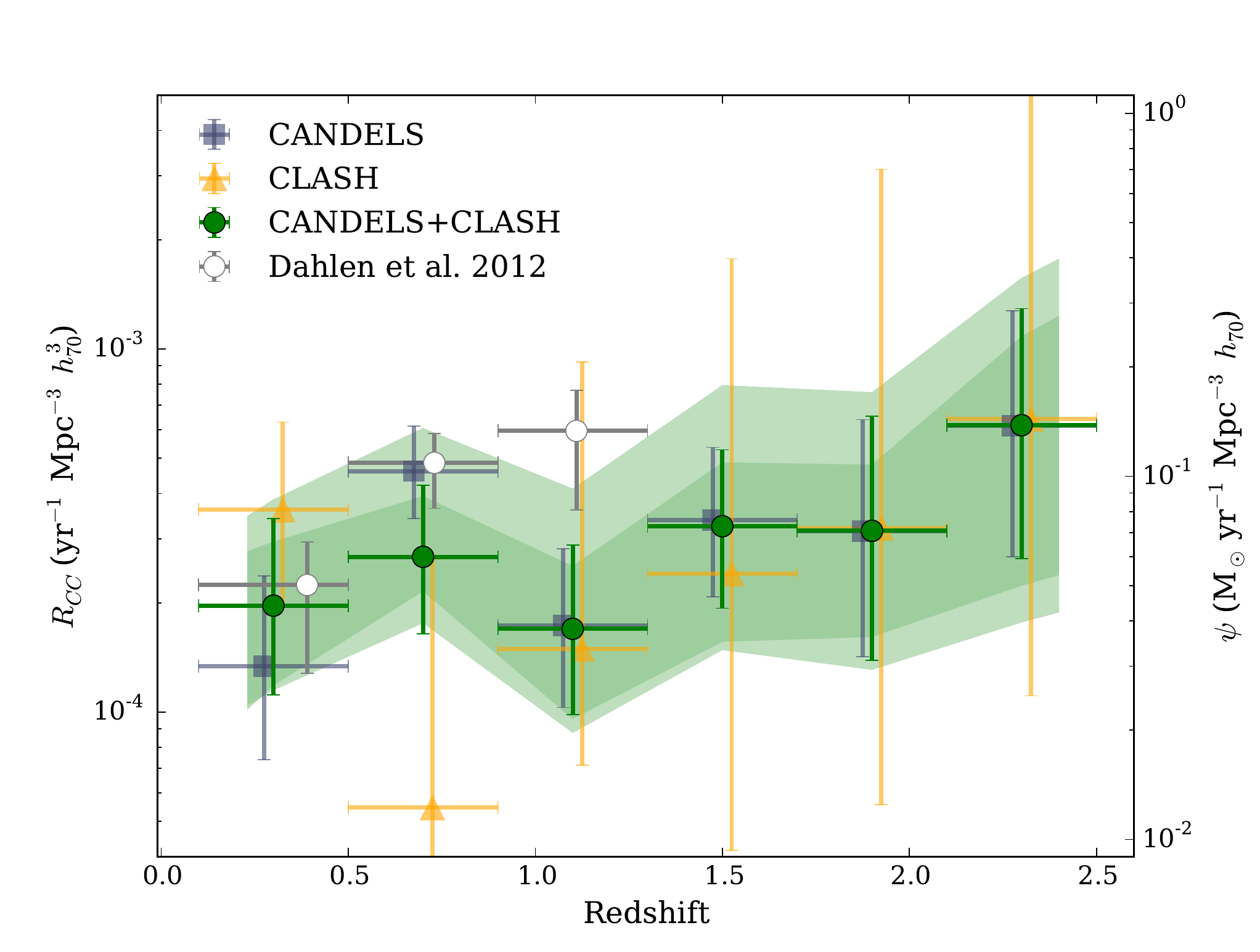}
	\caption{\footnotesize Volumetric core-collapse supernova rates from CANDELS (blue squares) and CLASH (yellow triangles), and the weight-average of both surveys (green circles). Points are offset horizontally for clarity. CCSNe rates from \citet[ white circles]{Dahlen:2012} are shown for comparison. Vertical error bars are the statistical uncertainties for each bin, while the horizontal error bars represent the width of the redshift bin. The green shaded region shows the systematic error range for the combined rates under two extinction assumptions: with variance in observed extinction, and under extreme limit scenarios discussed in Section~\ref{sec:results}. The right ordinate shows the scaled star-formation rate density, assuming $\psi=R_{CC}/[k\,h^2]$ with $k=0.0091\,M_{\odot}^{-1}$.}
	\label{fig:rates}
\end{figure*}

The CCSNe rates from CANDELS and CLASH, and the weighted average of both, are shown in six redshift bins from $z=0.1$ to $z=2.5$ in Table~\ref{tab:rates} and Figure~\ref{fig:rates}, with Poisson errors based on the observed number of events, derived from \cite{Gehrels:1986}. The systematic error range is shown in green, with two extinction scenarios as discussed below. For comparison, we  show the rates from our D12 analysis, with nearly the same number of events per redshift bin as are in the CANDELS+CLASH to $z<1.3$. 

\subsection{Systematic Errors in CANDELS+CLASH Rates}
Our approach to classification combines the traditionally systematic type and redshift uncertainties with the statistical error, making for a more complete, and perhaps fairer, assessment of overall rate uncertainty. However there are other factors which clearly contribute to the systematic assumptions in our calculation. Our additional systematic error budget is shown in Table~\ref{tab:errors} for each redshift bin, and includes the following factors:

\renewcommand{\arraystretch}{1.95}
\begin{deluxetable*}{lrrrrrr}
\tabletypesize{\footnotesize}
\tablecolumns{7}
\tablewidth{0pt}
\tablecaption{\sc Systematic Error Budget in CANDELS+CLASH Rates\tablenotemark{a}}
\tablehead{\colhead{Source}&\colhead{$\langle z \rangle = 0.3$}&\colhead{$\langle z \rangle = 0.7$}&
\colhead{$\langle z \rangle = 1.1$}&\colhead{$\langle z \rangle = 1.5$}&\colhead{$\langle z \rangle = 1.9$}&
\colhead{$\langle z \rangle = 2.3$}\label{tab:errors}}
\startdata
Type Fractions & $^{+0.00}_{-0.34}$ & $^{+0.00}_{-0.24}$ & $^{+0.00}_{-0.35}$ & $^{+0.00}_{-0.79}$ & $^{+0.00}_{-0.76}$ & $^{+0.00}_{-2.05}$ \\
$M_B$ & $^{+0.64}_{-0.27}$ & $^{+1.10}_{-0.26}$ & $^{+0.81}_{-0.42}$ & $^{+1.60}_{-1.02}$ & $^{+1.63}_{-0.99}$ & $^{+4.64}_{-2.65}$ \\
K-correction & $^{+0.71}_{-0.40}$ & $^{+0.67}_{-0.00}$ & $^{+0.16}_{-0.00}$ & $^{+0.01}_{-0.01}$ & $^{+0.11}_{-0.11}$ & $^{+0.11}_{-0.11}$ \\
Extinction (Normal Galaxies) & $^{+0.00}_{-0.48}$ & $^{+0.00}_{-0.36}$ & $^{+0.00}_{-0.49}$ & $^{+0.00}_{-1.06}$ & $^{+0.00}_{-0.89}$ & $^{+0.00}_{-2.02}$ \\
Extinction Limits & $^{+1.61}_{-0.54}$ & $^{+3.14}_{-0.83}$ & $^{+2.29}_{-0.60}$ & $^{+4.42}_{-1.19}$ & $^{+4.13}_{-1.35}$ & $^{+8.31}_{-2.82}$ \\
Lensing Degradation & $^{+0.14}_{-0.00}$ & $^{+0.33}_{-0.00}$ & $^{+0.09}_{-0.00}$ & $^{+0.03}_{-0.00}$ & $^{+0.05}_{-0.00}$ & $^{+0.29}_{-0.00}$ \\
$\Lambda$CDM & $^{+0.19}_{-0.18}$ & $^{+0.16}_{-0.15}$ & $^{+0.13}_{-0.12}$ & $^{+0.24}_{-0.23}$ & $^{+0.21}_{-0.20}$ & $^{+0.49}_{-0.47}$ \\
\hline
Total Systematic (Normal Galaxies)& $^{+0.98}_{-0.78}$ & $^{+1.26}_{-0.53}$ & $^{+0.83}_{-0.74}$ & $`^{+1.62}_{-1.69}$ & $^{+1.65}_{-1.55}$ & $^{+4.67}_{-3.94}$\\
Total Systematic (Extinction Limits)& $^{+1.89}_{-0.82}$ & $^{+3.39}_{-0.92}$ & $^{+2.43}_{-0.82}$ & $^{+4.70}_{-1.77}$ & $^{+4.45}_{-1.85}$ & $^{+9.54}_{-4.40}$
\enddata
\tablenotetext{a}{\footnotesize In units yr$^{-1}$ Mpc$^{-3}$ 10$^{-4}$ $h_{70}^{3}$}
\end{deluxetable*}

\vspace{0.1in}
\noindent{\it Type Fractions}: While we have adopted the \cite{Li:2011a} volume limited CCSN type fractions for completeness arguments, the fractions from \cite{Richardson:2014fk} may also have a reasonably complete sampling of the CCSN diversity. Using the ASC-derived type fractions results in rates reduced by up to $20\%$. 

\vspace{0.1in}
\noindent{\it Peak Magnitudes:} The value of the mean peak absolute magnitudes has a large impact on the discovery rate by shifting luminosity functions brighter or fainter against the survey efficiency function. Using the standard error in the mean peak absolute magnitudes for each type from  \cite{Richardson:2014fk} leads to variances in the rates of as much as 45\% in the highest redshift bins.

\vspace{0.1in}
\noindent{\it K-correction:} The k-corrections are determined from mean template SEDs provided by SNANA. We estimate the systematic uncertainty in these k-corrections using the dispersion in these mean corrections. It is important to note, the SNANA template libraries do not extend below $\sim$3000{\AA} or beyond $1\mu$m, making extrapolations from these templates to the synthetically observed passband somewhat uncertain. This is particularly worrisome in the lowest redshifts of this survey, at $z\la0.4$.  It is difficult to estimate the amount of uncertainty in this spectral region, as there are surprisingly few NIR spectra with matched light curves for each main CCSN type in the literature. We use the error provided by the extrapolation of each template comprising the mean in the k-correction of each event (used to find $t_c$) over the passband widths. We find k-corrections add a 24\% uncertainty to the $\langle z \rangle=0.3$ bin, and less than a 14\% uncertainty in all higher-$z$ bins.

\vspace{0.1in}
\noindent{\it Extinction in Host Galaxies:} As stated in Section~\ref{sec:ext}, we consider the hosts of these supernovae to be largely normal galaxies, similar to the population of galaxies observed at low-$z$. The environmental extinction along the line of sight has similarly been assumed to be consistent with  low-$z$ CCSN observations, with $\lambda_V = 0.187$. If we instead adopt an exponential distribution from more theoretical treatment of extinction, i.e., the MCMC predictions from~\cite{Riello:2005} on the randomized extinction of events in spiral galaxies at different viewing angles (and setting $\lambda_V=0.128$), it would result in rates reduced by up to 20\%, as shown in Table~\ref{tab:errors}. Our approach to estimating the host extinction in the synthetic rest frame of the observed filter by scaling from the $A_V$ using the Calzetti Law means that changes to the extinction law, in $R_V$ to within $\pm2$, has a negligible effect on rate values. 

However, nature could be hiding a large fraction of CCSNe from us. By virtue of being tied to star-forming regions, CCSNe are also likely to occur in the dustiest regions of galaxies. The strongest star-bursting galaxies, e.g. luminous and ultra-luminous IR galaxies (U/LIRGS), which should produce the most CCSNe, are also among the dustiest and UV/optically obscured~\citep{Mannucci:2007}. This juxtaposition makes it difficult to account for expected dimming or loss of events, and is a point equally important to understanding the discrepancy between UV and IR-derived star formation rates and rate density histories. 

To consider a viable range, we evaluate two additional extreme scenarios: a {\it no extinction limit}, where $P(A_V=0)=1$ for all model contexts in $t_c$; and a {\it high extinction scenario}, where a substantial fraction of high-$z$ hosts are U/LIRG-like, obscuring large fractions of SNe from sight. For the high-extinction scenario, we use a prescription provided by \cite{Mattila:2012} to account for the missing fractions by assessing SN budget in U/LIRGs, and tracking the contribution of these Arp 299-like galaxies to the whole as a function of redshift. We treat these high missing fractions as additional corrections to the rates derived for normal galaxies, using the highest statistical bounds as the `high extinction limit'. These limits are meant as true extrema, not likely representative of environments of most CCSNe. However, for completeness, we show these extrema in Figure~\ref{fig:rates} and Table~\ref{tab:rates}.

\vspace{0.1in}
\noindent{\it Lensing Degradation:} We have assumed an additional degradation in luminosity dispersions due to gravitational lensing. Assuming no lensing degradation, however, results in a rates increased by less than 10\%.

\vspace{0.1in}
\noindent{\it Cosmological Model:} Throughout this analysis we have assumed the concordance cosmology, with $\Omega_M=0.27$, $\Omega_\Lambda=0.73$, and $H_0=0.70$. Using the range of uncertainty from \cite{Rest:2014pd}, with $0.24 < \Omega_M < 0.3$ and $0.70 < \Omega_\Lambda < 0.76$, results in as much as a 6\% uncertainty in the resultant rates. 


\subsection{Comparisons to Other CCSN Rates}\label{sec:comparisons}
\begin{figure*}[t]
	\centering\includegraphics[width = 5.25in]{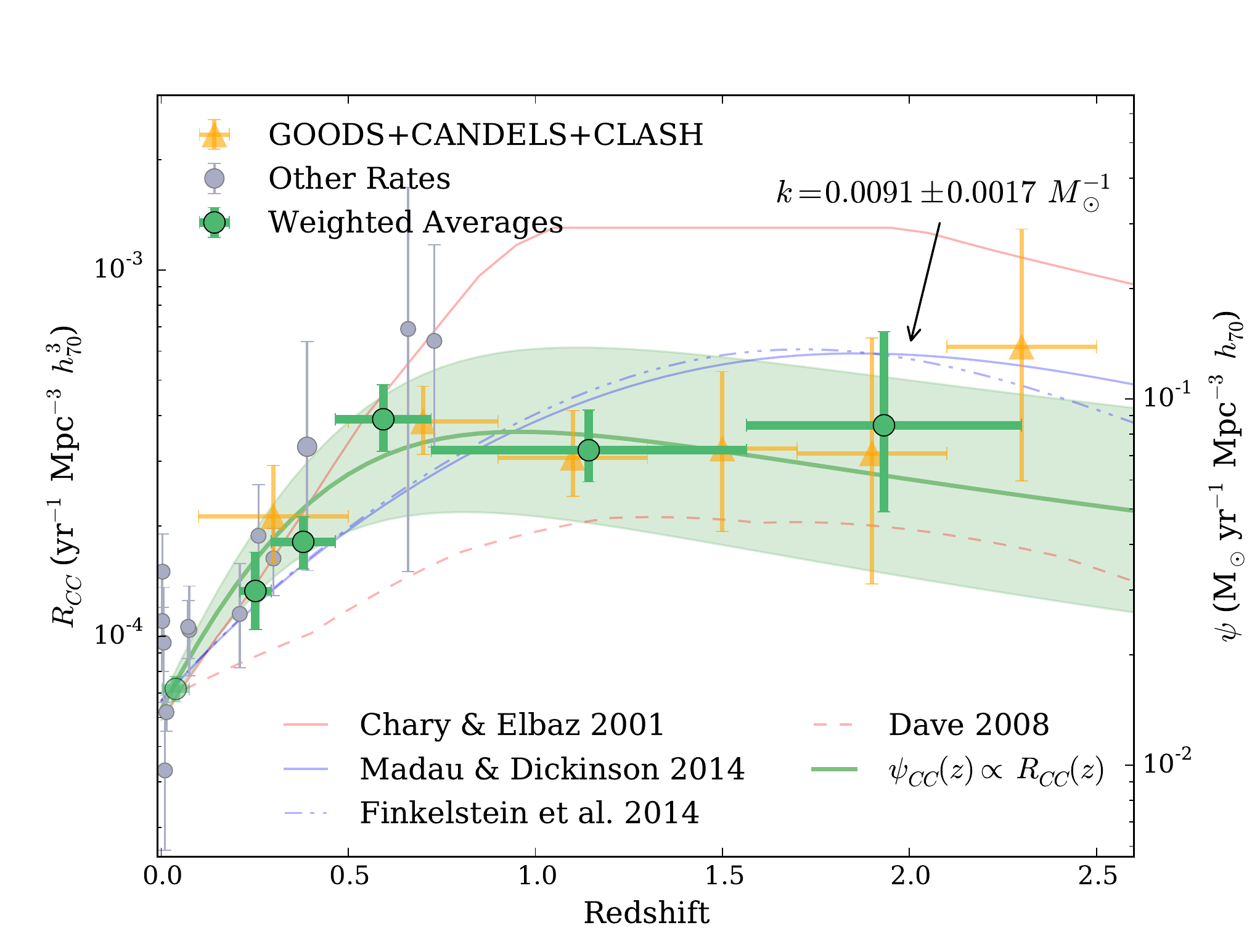}
	\caption{\footnotesize Rates from our group in comparison from other CCSN rates in the literature.{\it~Green circles:}~weighted average rates in six equalized redshift bins.{\it~Right Ordinate and Lines:} star-formation rate density models, scaled to best match the \citeauthor{Madau:2014fk} SFR to all CCSN rate measures, with $k=0.0091\pm0.0017\,M_{\odot}^{-1}$. Also shown is the SFR model derived from the CCSNe rates (green) using the \citeauthor{Madau:2014fk} parameterization.}
	\label{fig:rates2}
\end{figure*}

We combine the rates here with those from D12 to provide rates from the combined GOODS+CANDELS+CLASH programs (shown in Table~\ref{tab:rates2} and Figure~\ref{fig:rates2}), as they are similar in their instrumentation, detection thresholds, and other assumptions which go into their analysis. The reduced statistical uncertainty remains dominated by the sample uncertainty from this analysis, which could be further reduced with precise spectroscopic redshifts of the SN hosts, or a more consistent treatment of the D12 sample, in future studies.

Also in Figure~\ref{fig:rates2}, we compare our combined GOODS+CANDELS+CLASH rates to those from \citet[ $\langle z\rangle=0.21$]{Botticella:2008}, \citet[ $\langle z\rangle=0.26$ and converted from SNu]{Cappellaro:2005yu}, \citet[ $\langle z\rangle=0.3$]{Bazin:2009}, \citet[ $\langle z\rangle=0.39$ and $\langle z\rangle=0.73$]{Melinder:2012}, and \citet[ $\langle z\rangle=0.66$]{Graur:2011}, and to local $z<0.1$ rates from \citet{Cappellaro:1999},  \citet{Botticella:2008},  \citet{Smartt:2009}, \citet{Li:2011a}, \citet{Mattila:2012}, \cite{Taylor:2014rm}, and \cite{Graur:2015fk}. Assuming each is a valid measure of the CCSN rate at the mean or effective redshifts from each sample, whose certainty is characterized by statistical errors free from significant systematic offsets (i.e., ignoring systematic uncertainties), we determine a complete volumetric CCSN rate history, ${R_{CC}}(z)$, using weighted average rate measures in 5 equalized redshift bins in the range $0.21<z<2.34$, with one additional weighted average bin at $z=0.04\pm0.04$, as shown in Table~\ref{tab:rates2} and Figure~\ref{fig:rates2}.

\renewcommand{\arraystretch}{1}
\begin{deluxetable}{llr}
\tabletypesize{\footnotesize}
\tablecolumns{3}
\tablecaption{\sc Comprehensive CCSN Rates}
\tablehead{\colhead{Redshift}&\colhead{Rate\tablenotemark{a}}&\colhead{$N_{CC}$\tablenotemark{b}}\\
\label{tab:rates2}}
\startdata
\sidehead{{\it GOODS+CANDELS+CLASH:}}
0.3$\pm0.2$ & $2.13^{+0.80}_{-0.54}$ & 18.8\\
0.7$\pm0.2$ & $3.86^{+0.96}_{-0.72}$ & 40.6\\
1.1$\pm0.2$ & $3.07^{+1.06}_{-0.66}$ & 17.6\\
1.5$\pm0.2$ & $3.25^{+2.03}_{-1.32}$ & 6.4\\
1.9$\pm0.2$ & $3.16^{+3.37}_{-1.77}$ & 3.0\\
2.3$\pm0.2$ & $6.17^{+6.76}_{-3.52}$ & 2.7\\
\sidehead{${R_{CC}}(z)$\tablenotemark{c}:}
$0.04 \pm{0.04}$ & $0.72^{+0.06}_{-0.06}$&\nodata\\
$0.25 \pm{0.04}$ & $1.33^{+0.37}_{-0.29}$&\nodata\\
$0.38 \pm{0.09}$ & $1.81^{+0.31}_{-0.28}$&\nodata\\
$0.59 \pm{0.13}$ & $3.91^{+0.95}_{-0.71}$&\nodata\\
$1.14 \pm{0.42}$ & $3.22^{+0.93}_{-0.58}$&\nodata\\
$1.93 \pm{0.37}$ & $3.76^{+3.01}_{-1.58}$&\nodata\\
\enddata
\tablenotetext{a}{\footnotesize In units yr$^{-1}$ Mpc$^{-3}$ 10$^{-4}$ $h_{70}^{3}$ with statistical errors.}
\tablenotetext{b}{\footnotesize From Equation~\ref{eqn:cdf}, in the intervals specified in the redshift column.}
\tablenotetext{c}{\footnotesize Weighted averages of CCSNe rates here and in the literature. Uncertainties are standard errors in the weighted means.}
\end{deluxetable}

\subsection{Comparison to Cosmic Star Formation Rate Densities}
We compare our complete volumetric CCSN rate history to cosmic star formation rate densities, SFR or $\psi(z)$, in Figure~\ref{fig:rates2}. We show the consensus SFR derived from UV+IR galaxy measures, from \cite{Madau:2014fk} and \cite{Finkelstein:2014fj}, as well as a SFR derived from IR luminous galaxies \cite[e.g.,][]{Chary:2001}. The SFR, in units $M_\odot\, \rm{yr}^{-1}\, \rm{Mpc}^{-3}$, relates to the CCSN Rate, in units $\rm{yr}^{-1}\, \rm{Mpc}^{-3}$, by the fraction per unit mass of the initial mass function, IMF or $\xi(M)$, that is responsible for producing these CCSNe. We have determined this average scale empirically using a weighted least squares fit of the \citeauthor{Madau:2014fk} $\psi_{UV}(z)$ model to all CCSNe rate measures, $R_{CC}(z)$, assuming they are related by
\begin{equation}
R_{CC}(z)=k\,h^2\,\psi_{UV}(z).\label{eqn:rcc}
\end{equation}
With the scale, $k$, as the only free parameter, this comparison results in $k=0.0091\pm0.0017\,M_{\odot}^{-1}$, where uncertainties are estimated from the reduced statical errors and the extent of the high and no extinction corrected rate values for the CANDELS+CLASH data only. We evaluate the goodness of fit from the reduced $\chi^2$ of the model fit to all rate measures, and find $\chi^2_{\nu}=1.4$. A similar fit to the \cite{Chary:2001} $\psi_{IR}(z)$ model results in $k=0.006\pm0.002\,M_{\odot}^{-1}$, but the goodness-of-fit is poorer, with $\chi^2_{\nu}=5.3$. A KS-test on the same model yields a $D$-statistic of 0.42, with a two-tailed $p$-value$=0.05$, suggesting these distributions are unrelated to 95\% confidence.

In another comparison, we can attempt to predict the shape of $\psi(z)$ from the CCSN rates assuming the same parameterization as \cite{Madau:2014fk} for $\psi(z)$, such that
\begin{equation}
\psi(z) = \frac{A\,(1+z)^C}{((1+z)/B)^D+1}.\label{eqn:mdp}
\end{equation}
\noindent We fit the above function to all rate data using a Levenberg-Marquardt least-squares algorithm, resulting in the green line and error region shown in Figure~\ref{fig:rates2}, with $A=0.015\pm0.009$, $B=1.5\pm0.2$, $C=5.0\pm0.7$, and $D=6.1\pm0.5$. These values describe a SFR that rises to, and declines from, $z\approx 1$ more steeply than the\citeauthor{Madau:2014fk} model describes, and is likely to be inconsistent with SFR measures if extrapolated to very high-$z$. 

\subsection{Comparison of Initial Mass Functions}
As shown in D12, from a numerical assessment of $8-50\,M_{\odot}$ progenitor stars with a \cite{Salpeter:1955rw} IMF, where:
\begin{equation}
k = \frac{\int\limits_{8M_{\odot}}^{50M_{\odot}} \xi(M,z)\,dM}{\int\limits_{0.1M_{\odot}}^{125M_{\odot}} M\,\xi(M,z)\,dM},
\end{equation}
\noindent one expects $k=0.007^{+27\%}_{-31\%}\,M_{\odot}^{-1}$, which is 25\% lower than our fitted value but within the reasonable range of uncertainty, assuming the upper and lower mass bounds of CCSN progenitors remain in this range.

There has been considerable concern on the absence of CCSN progenitors in nearby galaxies at masses $>20\,M_{\odot}$, and that, perhaps, this mass range represents a rough upper limit to the intrinsic nature of CCSNe progenitors~\citep{Smartt:2015qy, Eldridge:2013kq}. If so, then the expected fraction of the IMF responsible for CCSNe would drop to $k=0.005\pm0.002 \,M_{\odot}^{-1}$ (Salpeter). This value does not appear to be supported by the data (to about the $2\sigma$ level), but is difficult to completely reject given the various caveats in dust corrections for both the SFR and $R_{CC}$. There is also considerable uncertainty in the nature of the IMF, with the expectations (given $8-50\,M_{\odot}$ progenitors) of $k=0.0104\,M_{\odot}^{-1}$  for a \cite{Kroupa:2001gf} IMF, or $k=0.0079\,M_{\odot}^{-1}$ for the steeper, high-mass ($\Gamma=1.45$) IMF of \cite{Weisz:2015fk}, both equally consistent with the observed rates. 

As a speculative solution to the $M_{\star}-$SFR discrepancy in galaxies, \cite{Dave:2008qy} suggested a model in which the IMF evolves with redshift to increasingly favor the production of higher mass stars in earlier cosmic epochs. In the \citeauthor{Dave:2008qy} model, the \citeauthor{Kroupa:2001gf} characteristic mass, $\hat{M}$, which marks the change in the IMF power-law slope for low-mass stars, moves to larger mass with redshift by
\begin{equation}
\hat{M}(z) = 0.5\,(1+z)^2\, M_{\odot}.\label{eqn:mz}
\end{equation}
\noindent The model argues that $\psi(z)$ models inferred from tracers of the light from predominantly high mass stars, assuming a present-day IMF, would overestimate the actual $\psi(z)$ at worsening rate with redshift. As the model suggests, scaling the observed $\psi_{UV}(z)$ by an evolving fraction of high-mass contributers might reveal a truer $\psi(z)$, one possibly traced by SN rates. We show the~\cite{Dave:2008qy} $\psi(z)$ model as the blue-dashed line in Figure~\ref{fig:rates2}. It is interesting to note, this implied $\psi(z)$ would be similar in form (at least over the considered redshift range) to that expected from corrections by the evolution in stellar mass density~\citep{Perez-Gonzalez:2008fj}, or perhaps from galaxy `downsizing'.

Testing this $\psi(z)$ as we have for the previous models, we find the best fit scaling to the $R_{CC}(z)$ data with $k=0.0102\,M_{\odot}^{-1}$. The $R_{CC}(z)$ distribution is, however, inconsistent with this $\psi(z)$ model, with a KS $D=0.58$,  rejecting the  \citeauthor{Dave:2008qy} evolving IMF $\psi(z)$ model at the $>99.99\%$ level.

\section{Summary}\label{sec:summary}
We have used a sample of approximately 44 CCSNe from the CANDELS and CLASH programs to extend volumetric CCSN rates to $z=2.5$. Taken together with our previous results, we show a history of the CCSN rates from {\it HST} alone extending over the wide redshift range of $0.1<z<2.5$, at a precision comparable to similar measures from the ground, at much lower $z$. In combination with literature rates, we present a comprehensive CCSN rate history, with statistical uncertainties at or below systematic uncertainties.

These composite rates are in good agreement with what would be expected from the consensus star-formation rate density history, assuming an average fraction of the IMF contributing to CCSNe progenitors of $k=0.0091\pm0.0017\,M_{\odot}^{-1}$. The data are inconsistent with SFR histories derived solely from IR luminous galaxies, and less supportive of a progenitor mass range confined to the range $8-20\, M_{\odot}$, assuming a Salpeter IMF.  The rate data also appear to rule out simple scenarios of an evolving (or relaxing) IMF.

We present a model SFR history derived from the CCSN rate data alone assuming a  \cite{Madau:2014fk} parameterization, which is somewhat consistent with the \citeauthor{Madau:2014fk} consensus $\psi(z)$ to $z\sim2.5$, but is likely to have difficultly matching other $\psi(z)$ at even higher-$z$.

\acknowledgments
This work is based on observations made with the NASA/ESA {\it Hubble Space Telescope}, delivered by the data archive team at the Space Telescope Science Institute, which is operated by the Association of Universities for Research in Astronomy, Inc. under NASA contract NAS 5-265555. These observations are associated with programs GO-12060 and GO-12099.

\facility{HST}

\appendix
\section{Cadence-Areas of the CANDELS and CLASH Surveys\label{sec:cadences}}
The full table of cadence-areas for each of the subfields in CANDELS, and all fields in CLASH, are shown in Tables~\ref{tab:fullcadences} and \ref{tab:fullcadences2}. Table~\ref{tab:fullcadences} lists the areas of each template-search image pair within each target field that are co-added if they share the same cadence, or time between visits to the same pointings. Table~\ref{tab:fullcadences2} lists similar combinations, but also combined for both parallel fields of all 25 CLASH target fields. See~\cite{Grogin:2011} for  a complete description of the CANDELS fields, and \cite{Rodney:2014fj} for a description of how the CANDELS SN survey varied from field to field. See  \cite{Postman:2012} for a complete description of the CLASH fields, and \cite{Graur:2014} for a description of how the CLASH SN survey varied from field to field. 
\renewcommand{\arraystretch}{1.0}
\begin{deluxetable*}{lcccccccccr}
\tabletypesize{\footnotesize}
\tablecolumns{11}
\tablewidth{0pt}
\tablecaption{\sc Cadence-Areas for the CANDELS~MCT Fields}

\tablehead{\colhead{Cadence}&\colhead{UDF}&\colhead{GSA-Deep}&\colhead{GSA-Wide}&\colhead{GNA-Deep}&\colhead{GNA-Wide SW}&\colhead{GNA-Wide NE}&\colhead{EGS}&\colhead{UDS}&\colhead{COS}&\colhead{Total}\\
\colhead{(days)}&\multicolumn{9}{c}{(sq. arcmin)}&\colhead{}\label{tab:fullcadences}}
\startdata
33&\nodata&\nodata&\nodata&\nodata&\nodata&\nodata&\nodata&\nodata&4.60&4.60\\
45&\nodata&4.54&\nodata&\nodata&\nodata&\nodata&\nodata&\nodata&\nodata&4.54\\
46&\nodata&44.08&4.16&\nodata&\nodata&\nodata&\nodata&\nodata&13.78&62.02\\
47&\nodata&29.94&\nodata&\nodata&\nodata&\nodata&\nodata&\nodata&78.12&108.06\\
48&\nodata&43.51&4.42&\nodata&4.73&\nodata&4.24&\nodata&72.79&129.69\\
49&\nodata&89.95&8.89&\nodata&4.49&\nodata&17.13&89.39&27.48&237.33\\
50&\nodata&65.13&13.04&15.05&4.56&\nodata&63.13&112.98&\nodata&273.89\\
51&\nodata&34.31&\nodata&56.60&3.27&\nodata&21.66&\nodata&\nodata&115.84\\
52&\nodata&106.88&\nodata&58.06&11.39&\nodata&\nodata&\nodata&\nodata&176.33\\
53&\nodata&59.88&\nodata&36.11&9.16&\nodata&\nodata&\nodata&\nodata&105.15\\
54&\nodata&24.87&\nodata&98.63&\nodata&\nodata&\nodata&\nodata&\nodata&123.50\\
55&\nodata&44.07&\nodata&100.71&\nodata&38.66&\nodata&\nodata&\nodata&183.44\\
56&\nodata&19.20&\nodata&78.08&3.76&\nodata&\nodata&\nodata&\nodata&101.04\\
57&\nodata&5.00&\nodata&84.33&\nodata&\nodata&\nodata&\nodata&\nodata&89.33\\
58&\nodata&\nodata&\nodata&24.66&\nodata&\nodata&\nodata&\nodata&\nodata&24.66\\
59&\nodata&\nodata&\nodata&69.20&\nodata&\nodata&\nodata&\nodata&\nodata&69.20\\
60&\nodata&\nodata&\nodata&56.43&\nodata&\nodata&\nodata&\nodata&\nodata&56.43\\
61&\nodata&5.00&\nodata&10.15&\nodata&\nodata&\nodata&\nodata&\nodata&15.15\\
62&\nodata&\nodata&\nodata&\nodata&\nodata&\nodata&\nodata&4.69&\nodata&4.69\\
65&5.11&\nodata&\nodata&4.99&\nodata&\nodata&\nodata&\nodata&\nodata&10.10\\
73&\nodata&4.40&\nodata&\nodata&\nodata&\nodata&\nodata&\nodata&\nodata&4.40\\
85&\nodata&\nodata&4.49&\nodata&\nodata&\nodata&\nodata&\nodata&\nodata&4.49\\
91&\nodata&24.94&\nodata&\nodata&\nodata&\nodata&\nodata&\nodata&\nodata&24.94\\
92&\nodata&5.02&\nodata&\nodata&\nodata&\nodata&\nodata&\nodata&\nodata&5.02\\
93&\nodata&34.84&\nodata&\nodata&\nodata&\nodata&\nodata&\nodata&\nodata&34.84\\
95&\nodata&4.81&\nodata&\nodata&\nodata&\nodata&\nodata&\nodata&\nodata&4.81\\
105&\nodata&4.89&\nodata&\nodata&\nodata&\nodata&\nodata&\nodata&\nodata&4.89\\
136&\nodata&4.99&\nodata&\nodata&\nodata&\nodata&\nodata&\nodata&\nodata&4.99\\
139&\nodata&\nodata&4.43&\nodata&\nodata&\nodata&\nodata&\nodata&\nodata&4.43\\
140&\nodata&\nodata&\nodata&4.87&\nodata&\nodata&\nodata&\nodata&\nodata&4.87\\
Sum=&5.11&660.25&39.43&697.87&41.36&38.66&106.16&207.06&196.77&1992.67
\enddata
\end{deluxetable*}

\begin{deluxetable}{lcc}
\tabletypesize{\footnotesize}
\tablecolumns{3}
\tablewidth{0pt}
\tablecaption{\sc Cadence-Areas for the CLASH~MCT Parallel Fields}

\tablehead{\colhead{Cadence}&\colhead{ACS/WFC}&\colhead{WFC3/IR}\\
	\colhead{(days)}&\multicolumn{2}{c}{(sq. arcmin)}\label{tab:fullcadences2}}
\startdata

6 & 4.81 & 11.14\\
7 & 4.81 & 11.14\\
8 & 4.81 & 11.14\\
10 & 9.62 & 22.28\\
11 & \nodata & 11.14\\
12 & 76.99 & 189.35\\
13 & 91.42 & 289.60\\
14 & 48.12 & 89.11\\
15 & 33.68 & 77.97\\
16 & 14.44 & 44.55\\
17 & 24.06 & 66.83\\
18 & 9.62 & 44.55\\
19 & 52.93 & 189.35\\
20 & 43.31 & 155.94\\
21 & 4.81 & 11.14\\
22 & 4.81 & \nodata\\
23 & 38.49 & 100.24\\
24 & 28.87 & 44.55\\
25 & 28.87 & 66.83\\
26 & 19.25 & 33.41\\
27 & 19.25 & 55.69\\
28 & 19.25 & 66.83\\
30 & 4.81 & 11.14\\
33 & 4.81 & \nodata\\
34 & 4.81 & \nodata\\
35 & \nodata & 11.14\\
39 & 19.25 & 22.28\\
40 & 9.62 & \nodata\\
41 & 4.81 & \nodata\\
84 & 4.81 & \nodata\\
Sum=& 634.92 & 1648.72
\enddata
\end{deluxetable}

\bibliography{strolger}{}

\end{document}